\documentclass[aps,prb,twocolumn]{revtex4}
\usepackage{graphicx}
\newcommand\beq{\begin{equation}}
\newcommand\eeq{\end{equation}}
\newcommand\bear{\begin{eqnarray}}
\newcommand\eear{\end{eqnarray}}

\begin{document}


\title{Current-Voltage Characteristics in Donor-Acceptor Systems :
Implications of a Spatially varying Electric Field
}

\author{ S. Lakshmi and Swapan K. Pati}
\affiliation{Theoretical Sciences Unit and Chemistry and Physics of Materials Unit,
Jawaharlal Nehru Center for Advanced Scientific Research,
Jakkur Campus, Bangalore 560 064, India.}

\date{\today}

\begin{abstract}

{We have studied the transport properties of a molecular device composed
of donor and acceptor moieties between two electrodes on either side.
The device is considered to be one-dimensional with different on-site energies 
and the non-equilibrium properties are calculated using Landauer's formalism. 
The current-voltage characteristics is found to be
asymmetric with a sharp Negative Differential Resistance at a 
critical bias on one side and very small current on the other side.
The NDR arises primarily due to the bias driven electronic structure change 
from one kind of insulating phase to another through a highly delocalized 
conducting phase.
Our model can be considered to be the simplest to explain the
experimental current-voltage characteristics observed in many molecular
devices.}
\end{abstract}
\maketitle

The study of electron transport through single molecules is gaining tremendous
attention in recent years owing to the wide variety of applications
that they can be used in\cite{rev}. Recent advances in experimental techniques
have allowed fabrication and measurement of current through such
nanoscale systems. Various molecular systems have already been 
demonstrated to behave
as wires, switches, diodes and RAMs\cite{reed,Cui,Kushmerick}. 
The ability of a molecule to switch between off and on states is one of its
most important applications. Experimentally, this has been observed in 
several organic molecules with various donor and acceptor substituents
\cite{Tour,Donhauser,exptNDR,Xiao}. 
Considerable amount of theoretical work based on semi-empirical to
{\it ab}-initio methods have also been performed to model molecular transport 
characteristics\cite{rev}.
Many explanations for this switching phenomenon, based on 
charging\cite{Seminario,2site}, reduction of the acceptor 
moiety \cite{Xiao}, twisting of the ring structure leading to conformational 
changes\cite{Taylor,Bredas,Ranjit,lakshmi}, bias driven changes in molecule-electrode
coupling\cite{Xshi}, have been proposed. Most of these require to impose some external factor
like the rotation of the middle ring etc in order for the external bias to
cause NDR at some bias. And most often they do not make a relation between structural 
preference and bias polarity and hence do not explain the asymmetry that
has been observed in the experiments. Also a comprehensive understanding 
of the switching phenomena in general cases is still elusive.

In this letter, we try to understand the reasons for the observed 
Negative Differential Resistance (NDR) in molecular wires based on 
a very simple donor-acceptor model. 
In fact, the asymmetry as well as NDR in forward bias that has
been observed in the Tour molecules\cite{reed,Tour} comes out naturally
out of this dimer model without having to invoke any external factors.
And the simplicity of this parametrized model makes it very tractable and 
lends physical insight into the factors causing the NDR.
We find that a spatially varying external forward bias switches the
electronic phases 
resulting in a sharp rise and fall in transmission through the device.
Various extensions of the dimer model together with different spatial variations
of the electrostatic potentials have also been considered.

A closer look at the structure of the Tour molecules (see inset of Fig.1) 
suggests that one part of the molecule with the donor group (NH$_2$) has
a positive on-site energy and the other with acceptor group (NO$_2$) has a negative 
on-site energy (Although this has been assumed here, we have later extended it
to describe systems with many localized segments in it, as is shown by 
quantum chemical calculations on the Tour molecules).
For such a two-level model involving a donor
and an acceptor, the Hamiltonian is given as: 
$H = \sum_{i=1,2} \epsilon_i a^{\dag}_{i}a_{i} +t(a^{\dag}_{1}a_{2}+hc)$,
where $\epsilon_i$ is the on-site energy of site $i$ and $t$ is the
hopping integral. This dimer is 
attached to two metallic electrodes on either side, which 
are assumed to be non-interacting semi-infinite 
one dimensional systems 
described by a tight-binding Hamiltonian with an energy bandwidth of $4\gamma$ where
$\gamma=10$eV\cite{mujica2}.
The Fermi energy of the electrodes is generally assumed to be a fitting parameter.
We assume that the Fermi energy lies halfway between these
equilibrium (zero bias) energies, though its location does not 
influence our main results. 

Spatial variation of bias on the structure is quite a central issue 
in this field\cite{Sdatta}. Except for a few cases 
(given later), we consider that it
drops on the device as a ramp function, varying linearly from one 
electrode to the other as:
$V_n=-nV/(N+1)$, where n is the site-index and N, the total number of sites. 
With the potential, the energies for the dimer can be written as:
\bear
E_{1,2}&=&\frac{\epsilon_1+\epsilon_2-V}{2} \nonumber \\
 &\mp&\frac{\sqrt{9(\epsilon_1-\epsilon_2)^2+36t^2+V^2+6V(\epsilon_1-\epsilon_2)}}{6}
\eear

\begin{figure}
\includegraphics[scale=0.3]{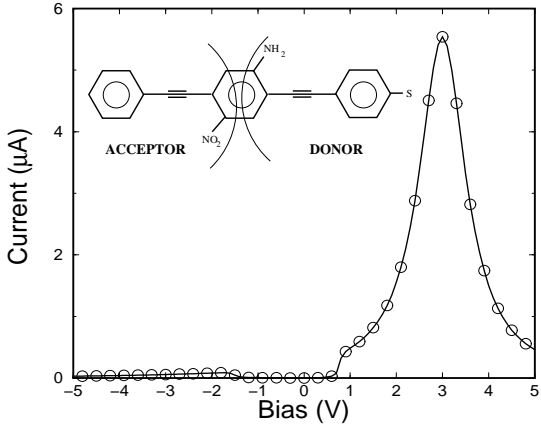}
\caption{The current-voltage characteristics for the 2-site system for
$\epsilon_2=-\epsilon_1=0.5 eV$ and $t=0.1 eV$. Inset is the Tour molecule.}
\end{figure}

\noindent The coupling to the electrodes modifies the bare Greens 
function of the molecule, which across the molecule can be written as
\bear
G_{12}&(E,V)&=\frac{V-3(\epsilon_2-\epsilon_1)-\sqrt{(3(\epsilon_2-\epsilon_1)-V)^2+36t^2}}{6t (E-E_1+i\Sigma_1+i\Sigma_2)} \nonumber \\
  &+&\frac{V-3(\epsilon_2-\epsilon_1)+\sqrt{(3(\epsilon_2-\epsilon_1)-V)^2+36t^2}}{6t (E-E_2+i\Sigma_1+i\Sigma_2)}
\eear

\noindent where $\Sigma_1$ and $\Sigma_2$ are the self-energies corresponding
to the two electrodes, calculated within the Newns-Anderson model\cite{Newns}.
Using the Greens function, the current through the system can
be obtained from the Landauer's formula \cite{datta_book}:
\beq 
I(V)={2e \over {h}}\int_{E_f-eV}^{E_f}dE \,
[Tr(\Gamma_1G\Gamma_2G^{\dag})]
\eeq
\noindent where $\Gamma_{1,2}$ are the anti-hermitian parts of the
self-energy matrices, $\Gamma_{1,2}=i(\Sigma_{1,2}-\Sigma^{\dag}_{1,2})$
\noindent which describe the broadening of the energy levels due to
the coupling to the electrodes. 

In zero bias condition, as can be seen from Eq.1 (with $V=0V$), presence of different 
on-site energies opens up a gap larger than that
for a purely hopping model ($\epsilon_2=\epsilon_1=0 eV$)
near the zero of energy indicating the preference of the electrons to stay 
at the atomic site with negative on-site energy. The equilibrium transmission 
is found to be large for purely hopping
model since it corresponds to equal distribution of charges.  
With the inclusion of different on-site energies, the systems becomes
insulating due to charge transfer and the zero-bias transmission reduces due to this
preferential charge localization.

\begin{figure}
\includegraphics[scale=0.4]{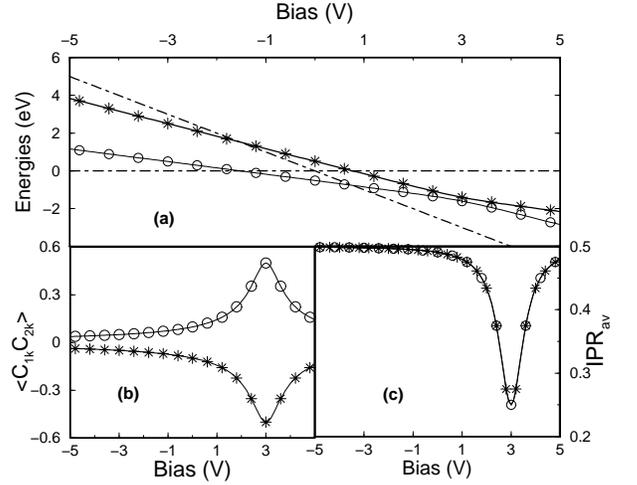}
\caption{(a) The variation of the two levels (circles and stars) 
of the 2-site system with the applied bias, for $\epsilon_2=-\epsilon_1=0.5 eV, t=0.1 eV$
The dotted lines indicate the variation
of the Fermi energies of the electrodes with bias.
(b) The numerator of the Greens function matrix element for the corresponding
energy levels shown in (a). 1 and 2 represent the site index and k specifies
the corresponding level.
(c) The $IPR_{av}$ for the levels shown in (a). See the text for definition.}
\end{figure}

Fig 1 shows the nature of the current-voltage characteristics with
external bias. As can be seen, the
current is negligible around the zero of energy and around a bias of $1 V$, 
there is 
a small jump in the current. This is the bias at which the device energy
level comes into resonance with the Fermi energy of the electrodes\cite{mujica2}.
With increase in the forward bias, 
around a bias of $3 V$, the current shows a 
sharp rise and fall,
indicating strong Negative Differential Resistance (NDR). On the other 
hand, with increase in the reverse bias, the system continues to 
remain insulating with negligible current.

To understand the reasons for the NDR, we look at the variation 
of the energy levels ($E_k$) of the bare molecular dimer with bias (Fig 2a) and
the numerator of the Greens function, $\langle 1|k \rangle \langle 2|k \rangle$ 
(see Fig 2b, $k=1,2$ are the eigenstates). With increase 
in the forward bias, the energy levels 
come close to one another up to
the critical bias $V_{c}$ at which the NDR is seen, above which, they 
move farther away. In Fig. 2b, exactly around this $V_{c}$, the
contribution to the eigenstate (MO) coefficients from the sites increases quite sharply.
To quantify this critical bias $V_{c}$,
we minimize the gap between the energies with respect to the applied bias and obtain,
$V_{c}=3(\epsilon_2-\epsilon_1)$, which is in accordance with our numerical data
(for $\epsilon_1=-0.5 eV$ and $\epsilon_2=0.5 eV$, we find $V_{c} \sim 3 V$) \cite{Pearson}.
At this critical bias, the energies take the values $\mp t$, precisely the
energies of the non-interacting system. 
However, with increase in the reverse bias, the energy levels start 
diverging away from their
zero bias gap making the system more and more insulating, explaining 
the small current that is observed in Fig.1 for negative bias.

To quantify the effect at $V_c$, we calculate the Average Inverse 
Participation Ratio ($IPR_{av}$) which defines the extent of localization for 
a given eigenstate, with $E_k$:
\beq
IPR_{av} = \frac{1}{D(E)}\frac{1}{N}\sum_{k} P_k^{-1}\delta(E-E_k)
\eeq

\noindent where $P_k^{-1}$ is the IPR, defined as 
$P_k^{-1}=\frac{1}{N}\sum_{j}|{\psi(j,k)}|^4$ where the $j$ is
the atomic site index and $D(E)$ is the density of states. 
Fig. 2c shows a strong dip in the values of $IPR_{av}$
around the critical bias indicating complete delocalization in the system, while
at other values of the bias, $IPR_{av}$ is much larger due to the localized
nature of the eigenstates.

\begin{figure}
\includegraphics[scale=0.4]{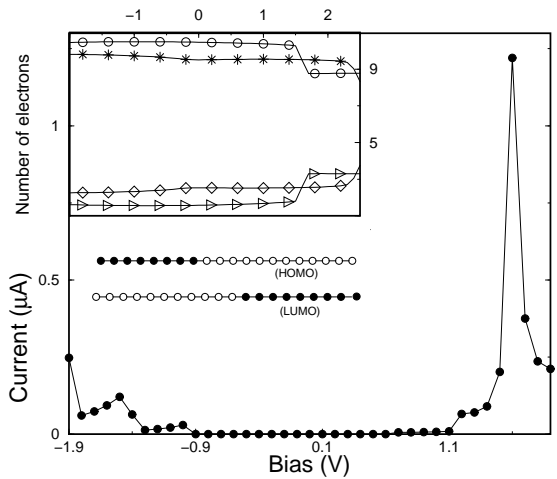}
\caption{The I-V characteristics of a 20 sites system, for 
$\epsilon_2=-\epsilon_1=0.5 eV$ and $t=1.0 eV$. The inset shows the variation of
the number density at the even sub-lattice for the case $i$ (triangles) and 
case $ii$
(diamonds) discussed in the text. Similarly in the odd sub-lattice for case $i$
(circles) and case $ii$ (stars). Also shown the pictorial representations
of the corresponding eigenfunctions in site basis, very close to the critical bias. 
Filled circles indicate large contribution to the eigenstate.}
\end{figure}


Initially for small bias, as noted before, the system tends to accumulate 
its charge density at the site with lower
on-site energy. Such a localization causes the system to be insulating. If this site 
is closer to the electrode with higher chemical potential, charges tend to move 
towards the other site. At $V_c$, where the NDR is seen, the charge densities 
are equally distributed at both
sites with no preference of one site over another indicating a situation
where both the on-site energies are equal. Further increase of bias would 
localize the charges
on the other site resembling an insulating dimer with its on-site 
energies interchanged, precisely the case as with reverse bias.

To extend these dimer results further, we look at a chain of $N$ atoms with
alternating on-site energies, $\epsilon_{1}$ and $\epsilon_{2}$. 
The energies in the absence of external bias 
have the form:

\beq
E_j=\frac{1}{2}  [(\epsilon_1+\epsilon_2)\mp \sqrt{(\epsilon_1-\epsilon_2)^2+ 16t^2 \cos^2(\frac{\pi j}{N+1})}]
\eeq

\noindent where N is the number of sites and the '$-(+)$' signs hold for 
j=1,2....,$\frac{N}{2}$ (j=$\frac{N}{2}$+1,.....N). 
It is quite well-known that different diagonal terms (either random or alternating) 
localizes the eigenstates of a one-dimensional system. From a perturbative 
treatment, this localization can be made
quantitative as $1/\lambda=\log|W/t|$ \cite{mujica2}, 
where the localization length $\lambda$ depends on the width of the diagonal 
term $W$ and the hopping $t$. Since the system becomes highly delocalized 
at $V_c$, we can naively expect the width to tend to zero at this bias, giving
$V_c|_{t \rightarrow 0}=(\epsilon_2-\epsilon_1)(N+1)/(N-1)$, for a $N$ sites
chain.

For a 20 sites half-filled system, we obtain a $V_c$ of $1.6V$ for values of 
$\epsilon_{1}=-0.5 eV$
and $\epsilon_{2}=0.5 eV$ and $t=1 eV$, which is close to the value calculated from the
above expression ($\sim 1.1V$), with second-order corrections due to the hopping term
\cite{Pearson}.
Fig 3 shows the I-V characteristics for this system which is asymmetric and 
shows clear eigenvalue 
jumps for reverse bias\cite{ndr_reverse} and a sharp NDR peak at a 
critical forward bias. 

To quantify the effect of bias in the system, we
calculate the total charge densities in each of the two sub-lattices 
for two cases: (i) charges are filled up to the highest-occupied level (HOMO) and 
(ii) with one of the highest occupied level electron promoted to the lowest
unoccupied level(LUMO). Both of these, for each sub-lattice are plotted in the inset 
of Fig.3, as a function of external bias. As can be seen, there is a jump 
in the charge density in the $+\epsilon$ sub-lattice due to the initiation 
of charge transfer from $-\epsilon$ 
sub-lattice to the sub-lattice with $+\epsilon$, exactly at the critical bias.
Since the HOMO and LUMO levels are
very close in energy at the critical bias (as 
shown for the 2-level system in Fig.2),
the charge densities corresponding to both cases become equal for each 
sub-lattices, giving rise to the crossing of the respective curves.
We have also shown in the figure, a pictorial representation 
of the HOMO and LUMO eigenstates close to $V_c$. As can be seen, 
the contribution to HOMO primarily comes from one end of the chain and to LUMO from 
the other end. However, since at $V_c$ these two states are quasi-degenerate,
a linear combination of the two correspond 
to a system with large transmission amplitude between two ends.


\begin{figure}[h]
\includegraphics[scale=0.4]{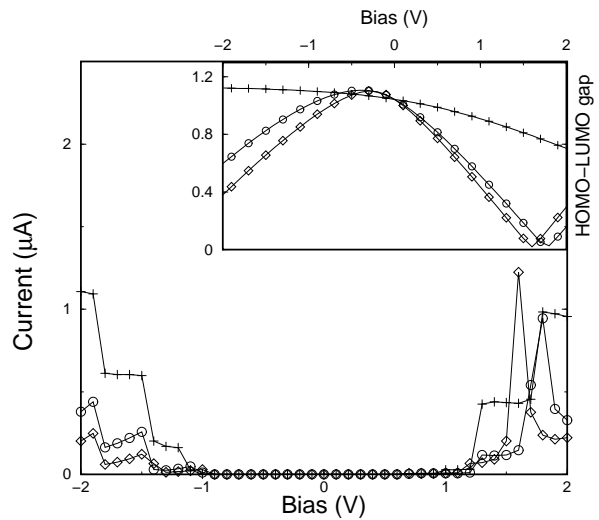}
\caption{ The I-V characteristics for the 20 site system with 
$\epsilon_2=-\epsilon_1=0.5 eV$ and $t=1 eV$ with ramp potential 
(diamonds), l=2 drop(plus), and l=8 drop (circles) close to the
interface. The inset shows the variation of the HOMO-LUMO gap
with bias for the three cases.}
\end{figure}

As is well-known, due to screening effects, the potential gradient is 
expected to be larger near the electrodes than at the center of the chain 
where it may be zero or very small\cite{Sdatta}. We have considered
the spatial variation of the potential over $l$ sites close to the interface
with variation in $l$, and the potential is a ramp function 
when $l=N$.
Three different potential drops have been considered: 
(a) ramp (b) $l=2$ (2 sites drop) and (c) $l=8$ ( 8 sites drop)
close to the interface. In Fig 4., we have plotted the I-V
characteristics for the system with $\epsilon_2=-\epsilon_1=0.5 eV$ and
$t=1.0 eV$ for all the three cases. For the one with $l=2$, the I-V curves 
shows clear eigenvalue jumps. With the increase in $l$, the NDR begins 
to appear at smaller voltages, and
a very sharp NDR peak is seen for the ramp potential.
The inset shows the variation of the HOMO-LUMO gap with bias 
for the three cases. There is no significant closing in
of the levels when $l=2$. But with increasing $l$, this becomes
more and more pronounced and the NDR appears earlier and 
is sharper. This study indicates the importance of the spatial
variation of the applied electric field in order to see an NDR response.

In the long-chain limit, there is an implicit relation between variation
in on-site energies and the bond-length alternation (BLA). Almost all the Tour 
molecules have BLA together with donor and acceptor groups. In fact, 
the BLA which dimerizes the system have been shown to give rise to NDR\cite{lakshmi}.
Interestingly, we find that donor ($+\epsilon$) and acceptor ($-\epsilon$) 
at some positions together with explicit dimerization also
causes NDR and asymmetric I-V, very similar to the I-V shown in Fig.3. 
The main point is that whether it is the 
explicit dimerization or two-sublattice structure, coupled with the voltage
drop, it induces interchange of symmetry together with Landau quasi-degeneracy of 
the low-lying levels.


To summarize, we find that the ratio 
$t : \epsilon_2-\epsilon_1$ is very crucial in determining the 
nature of the I-V characteristics.
Three features, namely, the critical bias, the sharpness of the 
NDR peak and the extent of asymmetry in the I-V curves are sensitive to this ratio. 
We believe this to be the reason for experimentally observed NDR and 
asymmetry in Tour molecules containing NH$_2$ and NO$_2$ groups or NO$_2$ group 
only, and its absence when there is only NH$_2$ group or no substituents.
Nitro group, having a very strong acceptor character tilts this
ratio in favor of a sharp NDR peak and asymmetry, within
the bias range considered.

Although our model neglects some of the important issues
in this field like electron-electron interactions,
Coulomb blockade, and other many-body effects, it is
nevertheless interesting to see how some experimentally observed
features can be captured even by such simple models.
And a two-level model, although not well-supported by quantum
chemical calculations, is not too far-fetched a description
of the Tour molecules, in which the strength of the localized
donor/acceptor reduces progressively towards the chain ends.

Our explanation for NDR encompasses the bias 
driven conformational change occurring in the system and the reduction of 
the acceptor to donor, where the roles of the donor and acceptor groups 
have been interchanged, as well as
the bias driven changes in electrode-molecule coupling which would 
result in a change in the on-site energies of the components of 
the molecule. However, the common thread among
all the known factors governing NDR requires further attention.


SKP acknowledges the DST, Govt. of India for financial support.


\begin{thebibliography}{99}
\bibitem{rev} A. P. Alivisatos {\it et al},  Adv. Mater. {\bf 10}, 1297 (1998);
F. Zahid, M. Paulsson and S. Datta, in {\it Advanced Semiconductors and 
Organic Nano-Techniques}, ed. by H. Morkoc, Academic Press, 2003.
\bibitem{reed} M. A. Reed et al, Science {\bf 278}, 252 (1997); Appl. Phys. Lett. {\bf 78}, 3735 (2001).
\bibitem{Cui} X. D. Cui et al, Science {\bf 294}, 571 (2001).
\bibitem{Kushmerick} J. G. Kushmerick et al, Phys. Rev. Lett. {\bf 89}, 086802 (2002).
\bibitem{Tour} J. Chen et al, Science, {\bf 286}, 1550 (2001); Appl. Phys. Lett. {\bf 77}, 1224 (2000).
\bibitem{Donhauser} Z. J. Donhauser et al, Science, {\bf 292}, 2303 (2001).
\bibitem{exptNDR}I. Kratochvilova et al, J. Mater. Chem. {\bf 12}, 2927 (2002); 
N. P. Guisinger et al, Nano Lett. {\bf 4}, 55 (2004).
\bibitem{Xiao} X. Xiao et al, J. Am. Chem. Soc. {\bf 127}, 9235 (2005).
\bibitem{Seminario} J. M. Seminario, A. G. Zacarias and J. M. Tour, 
J. Am. Chem. Soc. {\bf 122}, 3015 (2000); J. M. Seminario, A. G. Zacarias and 
P. A. Derosa, J. Chem. Phys. {\bf 116}, 1671 (2002).
\bibitem{2site} J. E. Han and V. H. Crespi, Appl. Phys. Lett. {\bf 79}, 2829 (2001).
\bibitem{Taylor} J. Taylor, M. Brandbyge and K. Strokbro, Phys. Rev. B. 
{\bf 68}, 121101 (2003).
\bibitem{Bredas} Y. Karzazi, J. Cornil and J. L. Bredas, J. Am. Chem. Soc. {\bf 123},
10076 (2001).
\bibitem{Ranjit} R. Pati and S. P. Karna, Phys. Rev. B. {\bf 69}
155419 (2004).
\bibitem{lakshmi} S. Lakshmi and Swapan K. Pati, J. Chem. Phys. {\bf 121},
11998 (2004).
\bibitem{Xshi} X. Shi, X. Zheng, Z. Dai, Y. Wang and Z. Zeng, J. Phys. Chem. B.
{\bf 109}, 3334 (2005)
\bibitem{our_prb} Y. Karzazi, J. Cornil and J. L. Bredas, Nanotechnology {\bf 14},
165 (2003); S. Lakshmi, A. Datta and S. Pati, Phys. Rev. B{\bf 72}, 
045131 (2005). 
\bibitem{mujica2} V. Mujica, M. Kemp, A. E. Roitberg and M. A. Ratner, 
J. Chem. Phys. {\bf 104}, 7296 (1996). 
\bibitem{Newns} P. W. Anderson, Phys. Rev., {\bf 124}, 41 (1961);
D. M. Newns, Phys. Rev., {\bf 178}, 1123 (1969).
\bibitem{Sdatta} S. Datta et al, Phys. Rev. Lett. {\bf 79}, 2530 (1997); 
V. Mujica, A. E. Roitberg and M. A. Ratner, J. Chem. Phys. {\bf 112}, 6834 (2000); 
S. Pleutin, H. Grabert, G-L Ingold and A. Nitzan, J. Chem. Phys. {\bf 118}, 3756 (2003).
\bibitem{datta_book} S. Datta, {\it Electronic Transport in Mesoscopic Systems}
(Cambridge Univ. Press, New York, 1996).
\bibitem{Pearson} R. G. Pearson, Inorg. Chem. 27, 734 (1988); Using the values
of the absolute electronegativities and hardness, the ionization potential 
difference between amino and nitro group is calculated to be of the
order of 1.0eV. Critical bias calculated from this, gives a resonable comparison with
experiments.
\bibitem{ndr_reverse} We also see an NDR peak at reverse bias region (at $\sim -2.6V$
compared to $\sim 1.6V$ in forward bias). This is because in the reverse bias, the 
charge movement is opposed by
the extra hopping term which arises due to the even number of sites in the
chain. The difference between the critical bias in the positive and negative
direction is governed dominantly by $t$ and to a second order
by the difference in the $\epsilon$. Note that, NDR on both sides of bias has been
observed in an experiment with Tour molecule recently\cite{Xiao}.
\end{thebibliography}
\end{document}